\documentclass[]{emulateapj}

\def\figdir{./}

\def\figname#1{\figdir/#1}

\def\kms{\ensuremath{\;\mathrm{km}\,\mathrm{s}^{-1}}}
\def\LCDM{$\Lambda$CDM}

\def\ltsima{$\; \buildrel < \over \sim \;$}
\def\simlt{\lower.5ex\hbox{\ltsima}}
\def\gtsima{$\; \buildrel > \over \sim \;$}
\def\simgt{\lower.5ex\hbox{\gtsima}}

\def\hide#1{}

\begin{document}

\pagestyle{myheadings} \markright{DRAFT: \today\hfill} 

\title{Galaxy Cluster Bulk Flows and Collision Velocities in QUMOND}

\author{Harley Katz$^1$, Stacy McGaugh$^{1,2}$, Peter Teuben$^1$, G. W. Angus$^3$} 
\affil{$^1$Department of Astronomy, University of
Maryland, College Park, MD 20742, USA}
\affil{$^2$Department of Astronomy, Case Western Reserve University, Cleveland, OH 44106, USA}
\affil{$^3$Astrophysics, Cosmology $\&$ Gravity Centre, University of Cape Town, Private Bag X3, Rondebosch, 7700, South Africa}
\email{hkatz@astro.umd.edu}
\email{stacy.mcgaugh@case.edu}
\email{teuben@astro.umd.edu}
\email{angus.gz@gmail.com}

\begin{abstract}
We examine the formation of clusters of galaxies in numerical simulations of a QUMOND cosmogony with massive sterile neutrinos.  Clusters formed in these exploratory simulations develop higher velocities than those found in \LCDM\ simulations.  The bulk motions of clusters attain $\sim 1000 \kms$ by low redshift, comparable to observations whereas \LCDM\ simulated clusters tend to fall short.  Similarly, high pairwise velocities are common in cluster-cluster collisions like the Bullet cluster.  There is also a propensity for the most massive clusters to be larger in QUMOND and to appear earlier than in \LCDM, potentially providing an explanation for ``pink elephants'' like El Gordo.  However, it is not obvious that the cluster mass function can be recovered.

\end{abstract}

\section{Introduction}

The formation of large scale structure is well understood with linear perturbation theory in \LCDM.  This paradigm provides a compelling description of the emergence of the cosmic web with massive clusters of galaxies forming at the nodes of filaments.  Fits to the acoustic power spectrum at $z \approx 1000$ \citep{WMAP5, Planck} and to the galaxy power spectrum at $z \approx 0$ \citep{Tegmark04} yield strong and consistent constraints on a modest number of cosmological parameters.  These observations provide strong support for a universe dominated by non-baryonic cold dark matter and dark energy.

While generally successful on large scales, the \LCDM\ paradigm suffers a number of shortcomings on smaller, galactic scales.  The predicted mass function of dark matter halos is a steep power law that bears no resemblance to the relatively flat galaxy luminosity function.  It has become conventional to invoke baryonic feedback processes to explain the manifestly non-linear mapping between the two, but there remains no completely satisfactory explanation of this phenomenon.  Many more sub-halos should exist within the virial radius of the Milky Way than are observed in the form of dwarf galaxies.  Feedback models have some success in addressing this problem \citep{MiaMassimo}, but also lead to further problems such as the ``too big to fail'' problem \citep{MiaGhosts,2big2fail}.  There exists no positive evidence for the existence of the large numbers of completely dark sub-halos that should be present in \LCDM.

Another persistent problem is the lack of cuspy mass distributions predicted for the central regions of dark matter halos by structure formation simulations in \LCDM.  Searches for the expected signatures of cusps have identified many galaxies in which the prediction fails \citep{OhTHINGS,TrachTHINGS,KdNMM09,dBcusprev}.  Feedback may also play a role in reshaping the cores of dark matter halos, but simulations of these effects \citep{fabiocuspynomore} remain far from providing an explanation for the many aspects of this problem \citep{KdNS11,KdNK11}.  More disturbing on a philosophical level is the persistent success of MOND \citep{originalmond} in predicting the aspects of the data that are problematic for \LCDM\ \citep{LivRelmondreview}.

Here we explore the role MOND might play in structure formation in the quasi-linear formulation proposed by \citet{QUMOND}.  This formulation is more amenable to numerical simulation than previous realizations of the theory.  \citet{angus2011} have implemented a code capable of performing cosmological simulations \citep[see also][]{qumondsolver}.  

One challenge is that it is not yet clear what the appropriate background cosmology is in MOND.  The timing of the development of structure depends on the expansion history, which is not well specified theoretically as in standard cosmology.  However, many of the successful aspects of \LCDM, including fits to the acoustic power spectrum of the cosmic microwave background, can be reproduced with a model that includes an $11 eV$ sterile neutrino \citep{angus2009}.  We adopt this model as a starting point for our exploration.  While such sterile neutrinos are conceivable, their chief value at present is to enable these considerations, acting as a proxy for the gaps in our knowledge much as dark matter and dark energy do in conventional cosmology.

While MOND has been extremely successful at galactic scales, it has been relatively unsuccessful at larger scales.  The accelerations in galaxy clusters tend to be higher than the MOND acceleration constant $a_0$ suggesting the need for some unseen matter to explain their dynamics \citep{Aguirre, Sanders2003, Sanders2007, AFB08}.  However, in order to maintain the success of MOND at galactic scales, the missing mass component particles must have a free streaming length that is larger than the sizes of typical galaxies in order to not interfere with the dynamics of MOND on the galactic level.  An obvious choice from the standard model would be the active neutrino which at certain masses is known to have free streaming lengths of orders larger than the size of typical galaxies \citep{Sanders2003, Sanders2007}.  \citet{angus2009} proposed that the addition of an $11eV$ sterile neutrino into the standard MONDian model would satisfy many cosmological constrains, including the primordial angular power spectrum.   \citet{angus2011} showed that this scenario leads naturally high pairwise velocities of nearby clusters.  

There have been other attempts to resolve the issues with MOND on larger scales without the inclusion of any sort of hot dark matter.  Recently, \cite{Zhao2012} proposed "extented MOND" (EMOND), where the effective force law depends on not only the acceleration, but also the depth of the potential well.  This solution has had success resolving the missing mass problem at all scales, not just in galaxies.

\citet{kashlinsky} observed that the bulk flows of galaxy clusters tend to be much higher than predicted by linear structure growth theory in a $\Lambda$CDM Universe.  Larger scale measurements of \citet{kashlinsky2010} completed in the same fashion agree with this earlier data that large scale bulk flows tend to be an order of magnitude higher than what is predicted for $\Lambda$CDM.  Similarly, but on a smaller scale, observations of 1E0657-558 (the Bullet Cluster) demonstrate that the relative pairwise velocity of the host and satellite cluster may exceed what is reproducible in a $\Lambda$CDM Universe \citep{leeandk}.  While \LCDM\ has had a difficult time producing these high velocity clusters in simulations, \cite{Llin2009} has found that high-speed collisions between host and satellite clusters are much more common in MOND simulations than in \LCDM\ simulations.


Although observations of the Bullet Cluster seemingly provide ``direct empirical proof of the existence of dark matter", the unique properties of this matter cannot be determined directly from these observations.  The weak lensing arguments only serve to show that the center of masses of the two clusters appear to be weakly interacting, while the motion of the gas seems to be hindered \citep{clowe}.  This allows us to test whether the $11eV$ neutrino proposed by \citet{angus2009} would be sufficient to reproduce such a system in a MONDian Universe.  

Using the QUMOND code \citep{{angus2009},{angus2011}}, we run eight $11eV$ sterile neutrino simulations with $256^3$ particles in a $(512 Mpc$ $h^{-1})^3$ box.
We investigate the mass and peculiar motions of clusters formed in these simulations.  The results are intriguing in the context of a number of recent observations: the early appearance of very massive clusters \citep[``pink elephants'' such as El Gordo:][]{elgordo}, the remarkably large bulk velocities of clusters of galaxies \citep{kashlinsky}, and the uncomfortably high collision velocity of the Bullet Cluster \citep{leeandk}.  For comparison, we run the same analysis code with a few necessary changes on the publicly available MultiDark simulation \citep{multidarkdata}.  This allows us to directly compare results from our $11eV$ MOND simulations to those of a $\Lambda$CDM simulation.

This paper is organized as follows: In Section 2, we briefly discuss the simulations and parameters used within.  In Section 3, we analyze the evolution of structure formation in the models via the mass function.  In Section 4, we discuss the peculiar velocities of clusters in our simulations by measuring bulk flows and comparing the measurements to observational data.  In Section 5, we analyze the probability of finding a Bullet Cluster type system and finally in Section 6 we present our discussion and conclusions.   

\section{The Simulations}
The QUMOND code is a cosmological particle-mesh code that uses multigrid methods (see Numerical Recipes \S 19.6  and \cite{Llin2008} for an introduction) to solve Poisson's equation. The advantage of QUMOND over AQUAL \citep{BekMil1984} is that solving for the MOND potential is more straight-forward. In AQUAL, one must solve a Poisson-like equation ($\nabla\cdot[\mu(|\nabla\phi|/a_0)\nabla\phi]=4\pi G\rho$) whose discretization produces a relaxation equation to give the MOND potential in a given cell that depends on the MOND potential in many surrounding cells.  Thus, this method takes many iterations to solve given that it is so non-linear.

In QUMOND, instead of going directly from density to MOND potential through a long relaxation stage, one solves for the MOND potential in a series of steps.  One first solves the standard Poisson equation to find the Newtonian potential ($\nabla ^2\Phi_N=4\pi G\rho$).  Then, there is an intermediate step where one takes derivatives of the Newtonian potential, which, when combined with an interpolating function, gives a new source term for a new Poisson-like equation ($\nabla\cdot[\nu(\nabla\Phi_N/a_o)\nabla\Phi_N]$). One equates this to $\nabla^2\Phi$ (where $\Phi$ is the MOND potential) to relax to the MOND potential and hence move the particles accordingly.

An interpolating function $\nu (\nabla\Phi_N/a_o)$ must be specified, and here we implement the one used by \cite{FB2005} to fit the terminal velocity curve of the Milky Way and by \cite{McG2008} to fit the rotation curve of the Milky Way.

The initial conditions for the simulations assume the sterile neutrino-baryon perturbations evolve as they would in general relativity until the starting point of the simulations at a redshift $z\thicksim200$. This assumes that MOND is unimportant prior to $z\thicksim200$.  The additional ansatz we employ is that the universe expands according to the Friedmann equation of general relativity with the defined amounts of matter and dark energy given in Table 1.  This convenient ansatz may merely approximate the true underlying cosmology.

As stated, we have run eight $11eV$ sterile neutrino MOND simulations with $256^3$ particles with masses of $5.996\times10^{11}M_{\odot}h^{-1}$ in a $(512 Mpc$ $h^{-1})^3$ box utilizing initial conditions from the COSMICS package \citep{bertschinger}.  For comparison, we utilize the publicly available MultiDark simulation with $2048^3$ particles with masses of $8.63\times10^9M_{\odot}h^{-1}$ in a $(1000 Mpc$ $h^{-1})^3$ box \citep{multidarksim}.   Initial conditions for these simulations can be found in Table 1.

\begin{table}[htdp]
\centering
\caption{Simulation Initial Conditions}
\begin{tabular}{c c c c c c c}
Simulation & $\Omega_b$ & $\Omega_{dm}$ & $\Omega_{\nu}$ & $\Omega_{\Lambda}$ & $h$ & $n_s$ \\
\hline
$11 eV$ & 0.0443 & 0.0 & 0.2255 & 0.7302 & 0.72 & 0.96 \\
MultiDark & 0.0469 & 0.2231 & 0.0& 0.73 & 0.70 & 0.95  \\
\hline
\end{tabular}
\label{default}
\end{table}

We utilize the Amiga Halo Finder (AHF: \cite{AHF, AHF1}) to extract the clusters from the simulation.  The halo finder used for the MulitDark simulation uses the Bound Density Maximum algorithm to also isolate halos with edges defined where $\rho=200\rho_{crit}$. 

\section{Mass Function}

\begin{figure*}[t]
\epsscale{0.8}
\plotone{\figname{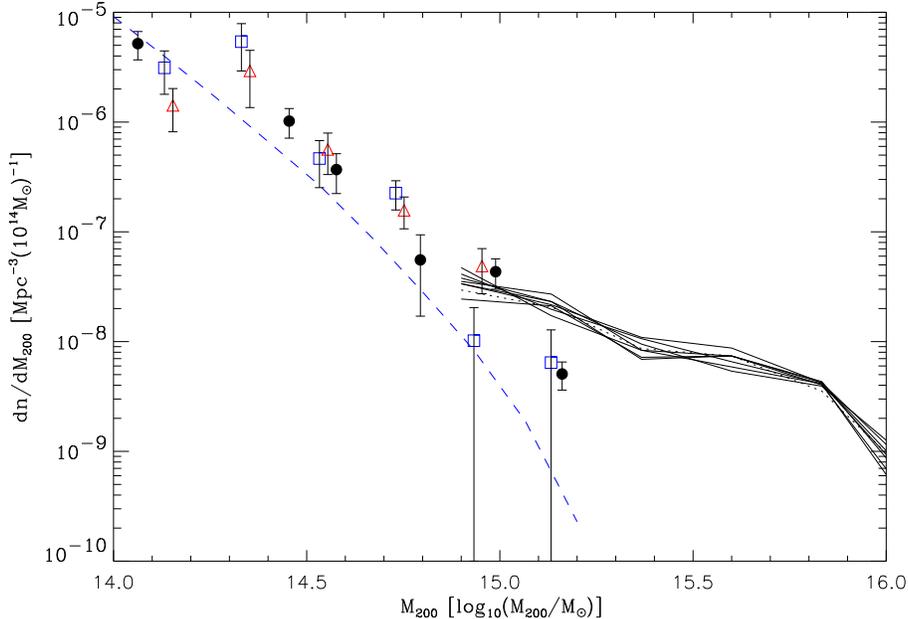}}
\caption{The red and black lines represent the $z=0$ cluster mass function for our eight $11eV$ MOND simulations.  The blue line shows the $z=0$ cluster mass function for the MultiDark simulation.  The data points are taken from \cite{MassFunc} where the filled circles are measured by the mass to $V_{max}(L_X)$ where the open points correspond to the $V_{max}(L(M_{tot}))$ relation.  The triangles and squares represent different $\alpha$ parameters in their $L_X$ to $M_{200}$.  For a more detailed description please refer to Section 4.2 of \cite{MassFunc}.}
\label{fig:}
\end{figure*}

Before quantifying how the motions of the clusters in our MOND simulations compare to observational data and a corresponding $\Lambda$CDM simulation, we seek to understand how the clusters themselves differ in the two cosmologies.  We utilize the mass catalog from the MultiDark database which measures the mass of a cluster out to its virial radius which is defined as where the enclosed density of particles reaches $200\rho_{crit}$.  This definition of cluster mass is conventional if rather arbitrary in $\Lambda$CDM.  We face a similar predicament in MOND.  It is rare that observations extend out to densities of $200\rho_{crit}$.  Most data truncate around $500\rho_{crit}$ which typically occurs around 1 Mpc \citep{VKFJMMvS}.  We adopt this as the radius at which we measure the masses of clusters in the QUMOND simulations.  The mass function from the eight $11eV$ MOND simulations is compared to the mass function from the MultiDark simulation as well as observational data in Figure 1.

\begin{figure*}[t]
\epsscale{.83}
\plotone{\figname{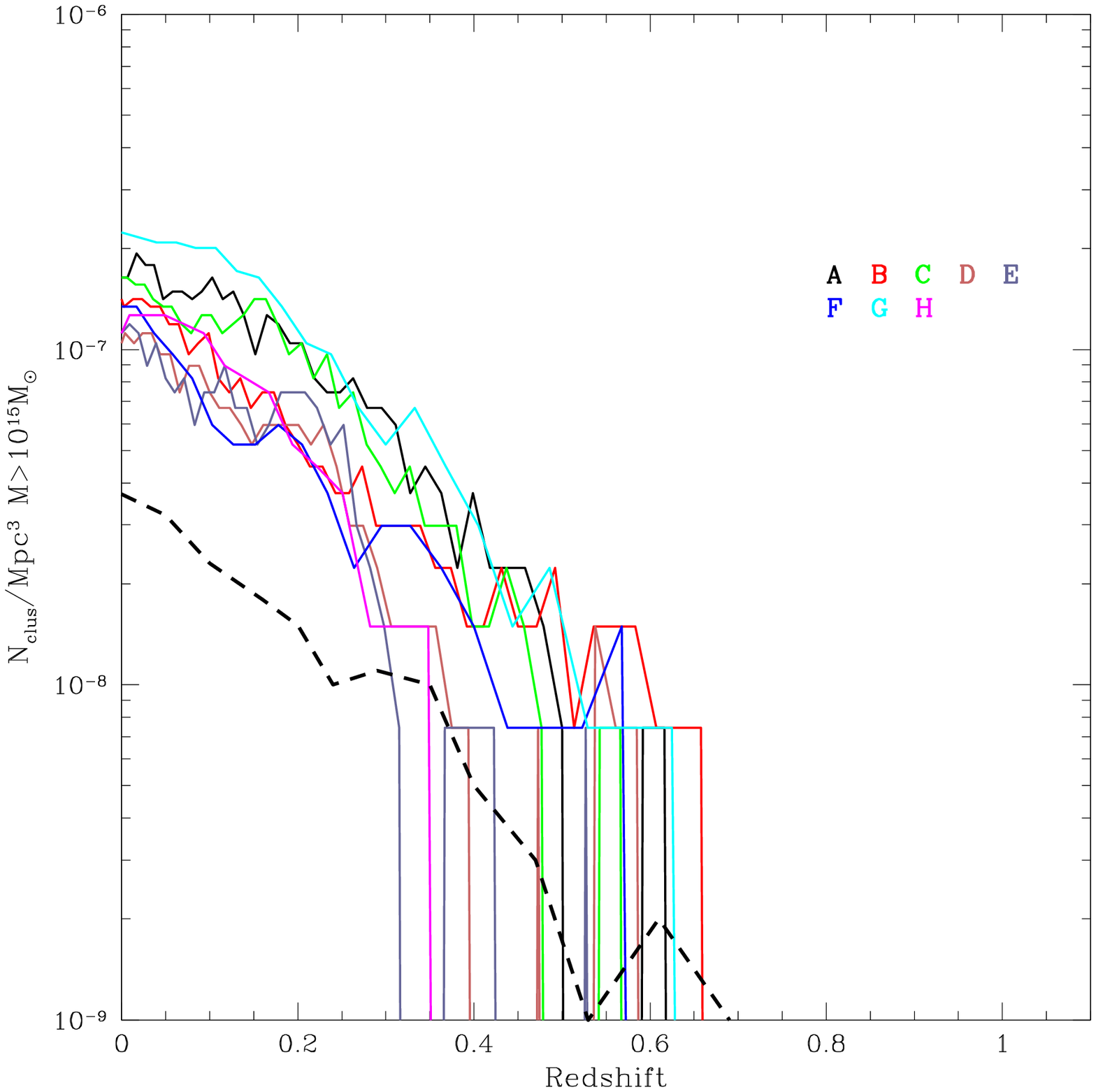}}
\caption{The dashed black line shows the density of clusters greater than $10^{15}M_{\odot}$ in the MultiDark Simulation.  The colored lines shows the density of clusters greater than $10^{15}M_{\odot}$ in our eight $11eV$ MOND simulations.}
\label{fig:}
\end{figure*}

At $z=0$, the total number of clusters identified by the halo finder in the MultiDark simulation was $1.3\times10^7$.  Accounting for the difference in the sizes of the simulations, we estimate that there are approximately $1.7\times10^6$ in any given $512Mpc/h$ cube.  Scaling once more by the resolution differences of the two simulations, we estimate there to be approximately 25,000 clusters in a $\Lambda$CDM simulation with the same box size and resolution as our own.  This does not include the differences in halo finders.  The average number of clusters identified by our halo finder in our simulations at $z=0$ is 190.  It is clear that the clusters formed in the $11eV$ MOND simulations are significantly less abundant than in the MultiDark simulation.  In Figure 2, we compare the number of $10^{15}M_{\odot}$ clusters in each simulation to the total volume of the simulation.  We see that these types of clusters are more rare in the $\Lambda$CDM simulation compared to the $11eV$ MOND simulations.  Because there are many more clusters in the MultiDark simulation, we can also infer that these clusters are at the higher end of the mass spectrum of the MultiDark simulation while they seem to typify an ordinary cluster in the $11eV$ MOND simulations.  This is consistent with the predictions of \cite{SandersNew}.

The conversion between MOND mass profile $M_m$ and the equivalent Newtonian mass profile $M_n$ is not a one to one relationship because of the difference in gravity profile (see eqn. 1) of \cite{angus2011}.  In Figure 2, as well as in all calculations in this paper, we have used the respective Newtonian masses in the simulation to compare the clusters from the two simulations, but it can be easily shown that the MOND mass is in fact smaller than the Newtonian mass at the radius we use to study the clusters.  At a radius of 1 $Mpc$, a $10^{14}M_{\odot}$ cluster would correspond to a Newtonian mass of $3.4\times10^{14}M_{\odot}$ \citep{angus2011}:
\begin{equation}
M_m=(M_n)^2\times(\frac{a_0r^2}{G} + M_n)^{-1}
\end{equation}
Just as we have found an over abundance of high mass clusters, \citet{angus2011}, taking into account this mass discrepancy, report a similar finding.

The existence of even one sufficiently massive high redshift galaxy cluster would falsify the $\Lambda$CDM model \citep{mortonson}.  The Bullet Cluster is of particular interest not only for its unique lensing properties, but also because it is one of the largest galaxy clusters currently observed at a redshift of $z\thicksim0.3$ with a mass of $\thicksim10^{15}h^{-1}M_{\odot}$.  Although this type of cluster is a rarity in $\Lambda$CDM cosmology, we find that such a massive cluster is of a typical size in our MOND simulations.  Figure 2 normalizes the number of high mass clusters in each of the simulations to their respective volumes.  It is clear that the probability of finding high mass clusters in a MONDian Universe is an order of magnitude higher than in a $\Lambda$CDM Universe.

The fair comparison of a cluster in a $\Lambda$CDM simulation to a cluster in one of the $11eV$ MOND simulations is far from obvious. The method we use throughout, where we take the mass of the cluster to be that within $1Mpc$, while simple, provides a reasonable method in which one might compare our simulated clusters to those from observations.  We want to emphasize that although a standard method for comparing MOND clusters to $\Lambda$CDM clusters has yet been developed, our results for every calculation other than the mass function remain independent of our definition of a cluster in MOND.  Methods to resolve this have been proposed in the works of \cite{Llin2009} and \cite{Knebe2009} where a dimensionless $V/\sigma$ relation can be used to compare clusters.  In Figure 3, we show the cumulative distribution function (CDF) for $V_{pec}/\sigma$ for two definitions of a cluster cutoff radius in our eight $11eV$ MOND simulations ($1Mpc$ and $R_{200}$) compared with that of the MultiDark simulation.  The CDF for the $11eV$ MOND clusters with a cutoff radius of $1Mpc$ is nearly identical to that of the MultiDark clusters which suggests that this might be a more fair comparison than using a cutoff radius of $R_{200}$.  Each method contains its unique drawbacks and therefore, we have avoided calculations (other than the Mass Function) where the results are sensitive to this definition.

\begin{figure}[t]
\epsscale{1.2}
\plotone{\figname{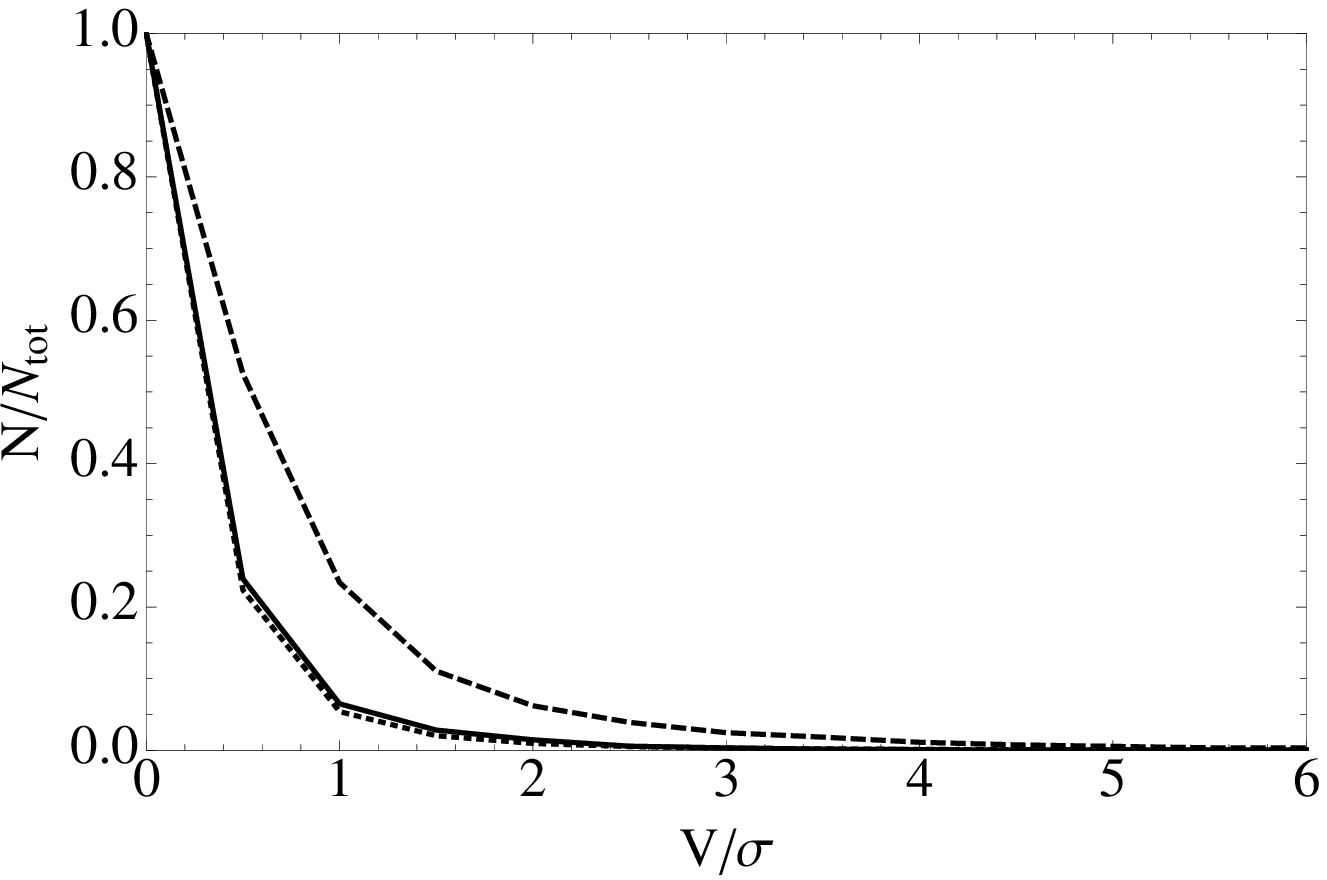}}
\caption{The dashed black line is the CDF for clusters in the eight $11eV$ MOND simulations with a cutoff radius of $R_{200}$.  The solid line is the CDF for clusters in the eight $11eV$ MOND simulations with a cutoff radius of $1Mpc$ and the dotted line is the CDF for clusters in the MultiDark simulation.  The $1Mpc$ cutoff radius clusters and the MultiDark clusters have a nearly identical CDF.}
\label{fig:}
\end{figure}

\section{Bulk Flow Measurements}
Multiple groups have argued that the probability of obtaining a pairwise velocity similar to that of the Bullet Cluster is extraordinarily low in $\Lambda$CDM cosmology (i.e. \cite{farrarandrosen, leeandk}).  However, this is not the only velocity measurement that has been found to be incompatible.  \citet{kashlinsky} demonstrated that the theoretical bulk flow measurements based on linear structure growth theory in $\Lambda$CDM cosmology are far smaller than observational measurements.  While these measurements often have large error bars because they are calculated indirectly from observations of the kinematic Sunyaev-Zel'dovich (kSZ) effect, the $\Lambda$CDM model still falls far short of the observational data.  Unfortunately, the growth of structure in a MONDian universe is not currently well understood; thus, we cannot make theoretical predictions on how we would expect the bulk flow measurements to behave based on a widely accepted structure formation law.  For this reason, we have attempted to numerically extract bulk flows from the $11eV$ MOND simulations.  Additionally, we attempt the same calculation on the MultiDark simulation to compare the numerical result with the theoretical predictions as well as the MOND data. 

Unfortunately, there remains no easy comparison between the bulk flows measured observationally and what can be extracted from numerical simulations.  The magnitude limits of the surveys prevent clusters under a certain flux from being observed.  This causes issues when comparing to our $11eV$ MOND simulations as well as the MultiDark simulation, because both simulations do not include baryonic matter.  Correcting for this bias becomes extremely difficult and we conservatively suggest that $10^{14}M_{\odot}$ clusters would be visible to the distances we probe in the numerical calculation for any of the surveys and thus use this mass as a threshold.

There have been multiple attempts to correct for many of the biases of the observational surveys; however, it remains uncertain which biases dominate for each survey considering they sample different regions of the sky.  \citet{watkins} attempt to match the bias of the surveys by creating a mock catalog which mimics the distribution of masses of clusters in the observations, and \citet{bouillot} assign a non-spherical bias parameter to account for the non-spherical distribution of the clusters in the survey.  Additionally, it is not clear if the biases that were corrected for in these attempts distort the bulk flow calculations to the same degree in MOND and $\Lambda$CDM cosmogonies.  

In a numerical simulation, we are fortunate to have the mass, position, and velocity evolution of every cluster in the simulation as a function of redshift.  This allows us to be able to extract the real $z=0$ bulk flows within any given volume.  However, as an observer at the Milky Way, no matter what technique is used to probe the peculiar velocity of another object, we are limited to knowing only the peculiar velocity at the observed redshift of that object.  Thus, simply averaging the peculiar motions of the galaxy clusters in a given volume within a numerical simulation is not an accurate technique to compare with observational data.  One must account for this redshift evolution, particularly in MOND simulations where the evolution of the velocity is much stronger than in $\Lambda$CDM.

\subsection{Velocity Function}

\begin{figure*}[t]
\epsscale{.8}
\plotone{\figname{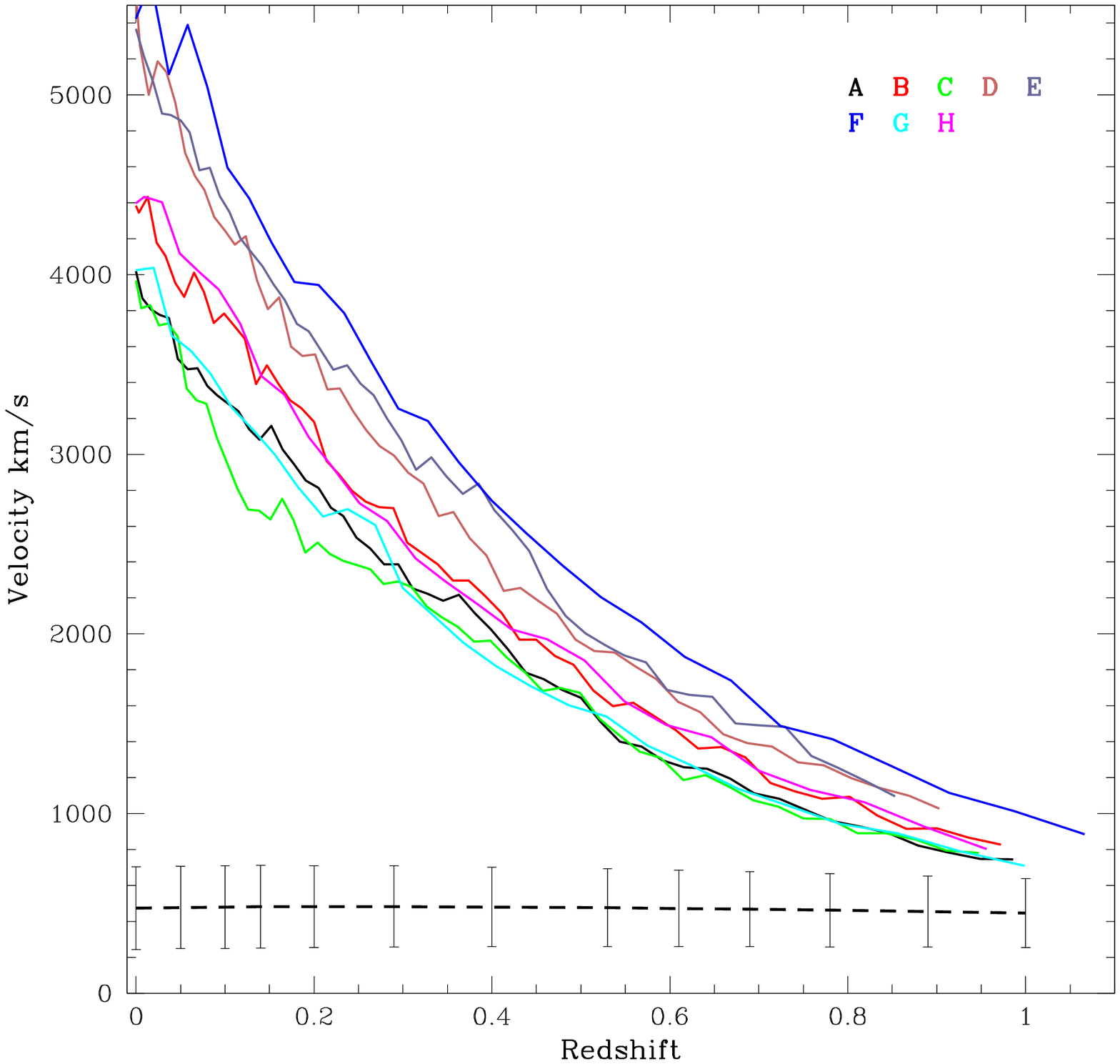}}
\caption{The dashed black line shows the median magnitude of the velocity of all of the clusters in the MultiDark Simulation along with 1$\sigma$ deviations.  The colored lines show the median magnitude velocity for all of the clusters in each of the eight $11eV$ MOND simulations.}
\label{fig:}
\end{figure*}

In order to measure the bulk flows with respect to any arbitrary cluster in the simulations, we must account for the evolution of each cluster's peculiar velocity as a function of redshift.  To achieve this we simply find the median of the magnitudes of cluster peculiar velocities at each redshift.  It must be noted that the peculiar velocity that we derive for a cluster is independent of the cluster cutoff radius.  We have purposefully left mass out of all of the following calculations so that our results would be independent of the mass we measure for each cluster and thus directly comparable to any such calculations done with a $\Lambda$CDM simulation.

In Figure 4, we show the MultiDark velocity function as well as the MOND velocity function.  While the MONDian velocity function appears as an exponential function and continually increases with decreasing redshift, the $\Lambda$CDM velocity function appears relatively flat.  In order to compare our bulk flow measurements to the observations of \citet{kashlinsky}, we need only be concerned with the regime later than $z\thicksim0.1$.  For this reason, we fit the first three data points of the MultiDark velocity function to the following least squares fit: $\overline{v}_{dm}=60.4z+473.6$.  Because the MONDian data points appear to follow a strict exponential throughout its evolution,  we fit the MOND simulations velocity function to the following exponential: $\overline{v}_{mond}=ae^{-bz}$ where the values for $a$ and $b$ for each of the eight simulations can be found in Table 2.  Errors on the $a$ values are $\sim 1\%$ and errors on the $b$ values are $\sim 0.5\%$.

\begin{table}[htdp]
\centering
\caption{MOND Velocity Functions}
\begin{tabular}{c c c }
Simulation & $a$ & $b$  \\
\hline
A & 3963 & 1.78 \\
B & 4407 & 1.79 \\
C & 3685 & 1.71 \\
D & 5166 & 1.88 \\
E & 5295 & 1.83 \\
F & 5578 & 1.75  \\
G & 3758 & 1.86 \\
H & 4428 & 1.79 \\
\hline
\end{tabular}
\label{default}
\end{table}

\subsection{Bulk Flow Calculation}
In order to determine bulk flows within the simulation, we employ the following formula:

\begin{equation}
v_{bulk}=||\frac{1}{\sum\limits_{n}^{N_{clus,r<R}}M_n}\sum\limits_{n}^{N_{clus,r<R}} \vec{v}_{n}(z)M_n||
\end{equation}
where $\vec{v}_{n}$ is the cluster's velocity at redshift $z_{n}$ with respect to an arbitrary initial cluster and $M_n$ is the mass of the cluster.  This $\vec{v}_{n}(z)$ is calculated according to the apparent redshift by:

\begin{equation}
\vec{v}_{n}(z)=\sum\limits_{p=\hat{i},\hat{j},\hat{k}}\frac{\pm ae^{-bz}}{\sqrt{3}}\pm|v_p(0)|-\frac{a}{\sqrt{3}}
\end{equation}
for the MOND simulations where $a$ and $b$ for each of the eight $11eV$ MOND simulations can be found in Table 2.  The first term simply sets the redshift dependence of the velocity as a function of redshift, and the second term shifts the function up and down so that the $z=0$ velocity matches that of the individual cluster.  Here we choose to calculate the mass weighted bulk flows within the simulations in order to compare with the observational data of \citet{kashlinsky}, which is biased towards the mass of the cluster observed from the use of the kSZ effect \citep{Li}.

\begin{figure*}[t]
\epsscale{0.73}
\plotone{\figname{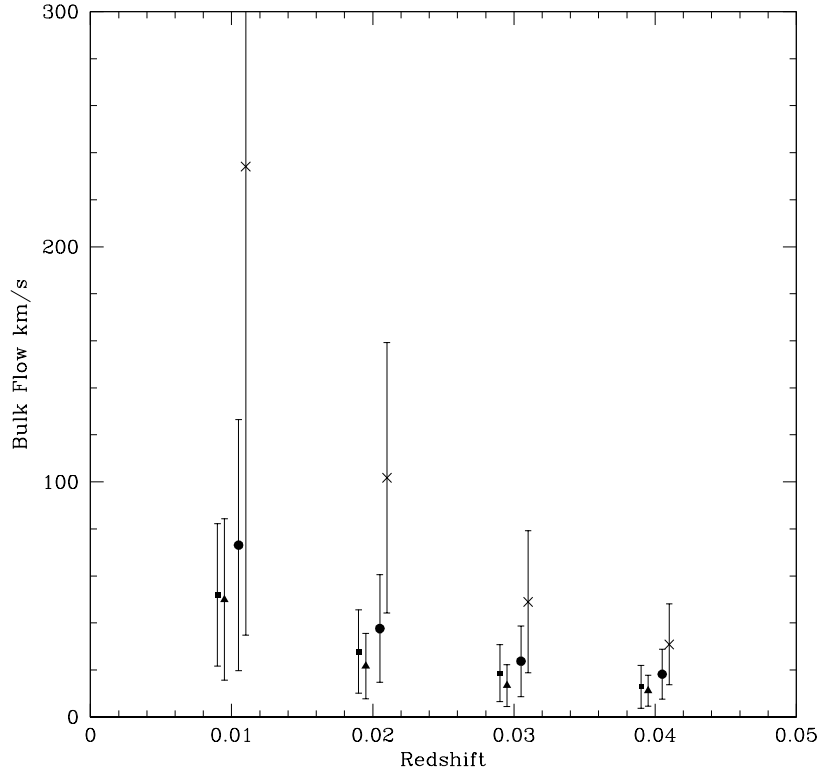}}
\caption{ The bulk flows measured as a function of redshift and mass cut for the MultiDark Simulation.  The square points have no mass cut.  The triangle points have a mass cut at $10^{12}M_{\odot}$.  The circle points have a mass cut of $10^{13}M_{\odot}$ and the cross points have a mass cut of $10^{14}M_{\odot}$.  All measurements were made at the same redshifts indicated on the axis and the dispersion is used to differentiate the points.}
\label{fig:}
\end{figure*}

However, the comparison between the MultiDark simulation and our MOND simulations remains inexact because of the differences in resolution.  Because all of the clusters in the MOND simulations are greater than $10^{14}M_{\odot}$, all clusters would be visible in surveys out to the depths that we measure in the simulations.  One might be concerned that the velocities could be different for the entire cluster compared to only the inner 1 Mpc cores which we are studying.  We have found that there remains no significant difference in the cluster velocity if larger radii are considered, because the overall cluster motion is consistent with the innermost densest region.  

To compare with the MultiDark simulation, we make four different mass cuts in the MultiDark simulation and compare the bulk flows (See Fig. 4).  These measurements do not take into account any velocity function because the velocity function changes at different mass cuts.  As the mass cut increases, the velocity function tends to flatten and become negligible.  The velocity function without any mass threshold is flat to within 1$\sigma$ errors and thus we do not include it in the measurements of Figure 5.  As the mass threshold increases, the measured bulk flows also tend to increase.  

We believe that the bulk flows are dominated by the larger structures, and the magnitudes of the bulk flows are suppressed by the smaller structures.  Additionally, the number of clusters within each sphere decreases with an increasing mass cut.  This may increase the measured bulk flows in the smaller spheres because objects within smaller distances tend to move in the same direction.  

It appears that the $10^{14}M_{\odot}$ mass cut might be the most appropriate to compare with the MOND simulations.  This limits the MultiDark simulation to clusters only the size of the MOND clusters and this is a conservative estimate to which clusters would be visible in observational surveys.  Figure 5 shows that the different mass cuts cause the greatest difference in the smallest sphere.  This difference is only on the order of $\thicksim175\;\kms$.  If we had a MOND simulation of equal resolution to the MultiDark simulation, we might find that the total bulk flow would decrease by a similar value.  This effect is negligible and within the error bars of the measured bulk flows in the MOND simulations.

In order to measure the empirical bulk flows within the MultiDark simulation we include the velocity function according to the apparent redshift as follows:

\begin{equation}
\vec{v}_{n}(z)=\sum\limits_{p=\hat{i},\hat{j},\hat{k}} \frac{\pm cz}{\sqrt{3}} \pm |\vec{v}_{p(0)}|
\end{equation}

where $c=60.4$.  It is clear that the redshift dependence of the peculiar velocities of clusters in the MultiDark simulation is significantly less than what is found in our $11eV$ MOND simulations.

\begin{figure*}[t]
\epsscale{0.7}
\plotone{\figname{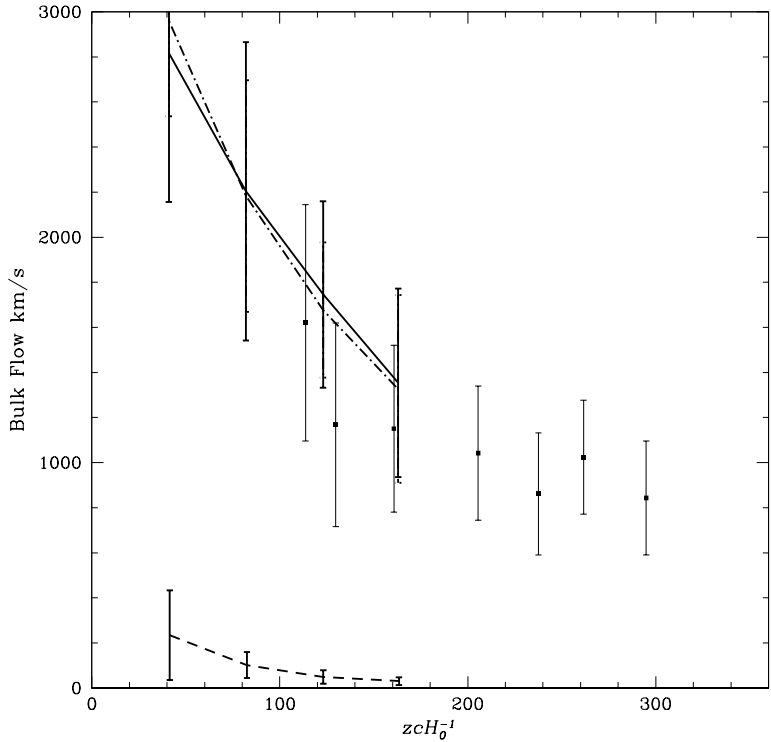}}
\caption{The dot dashed line shows the median bulk flows in each of the eight $11eV$ MOND where we define the cluster to be within 1Mpc.  The solid line shows the median bulk flows in each of the eight $11eV$ MOND where we define the cluster out to $r_{200}$.  It is clear that both definitions of a cluster result in the same computed bulk flow.  The individual square points are the observationally measured bulk flows from \citet{kashlinsky}.  The bottom dashed line data set is the $10^{14}M_{\odot}$ mass cut bulk flows from the MultiDark simulation.  The $1\sigma$ standard deviations are shown for the eight $11eV$ MOND simulations as well as the MultiDark simulation.}
\label{fig:}
\end{figure*}

To extract the bulk flows, we randomly choose a cluster in the simulation as an observer's reference frame.  We then measure the perceived redshift of each of the other clusters in the simulation with respect to the initial cluster and modify the peculiar velocity of each of the other clusters according to Equations 3 and 4.  We then find the mass weighted average of the peculiar velocities of the clusters.  We run 200 iterations of this calculation and find the median bulk flow.  For both the MultiDark and $11eV$ MOND simulations, we measure the net bulk flow of spheres of sizes corresponding to $z=0.01, 0.02, 0.03, 0.04$.  Because of the finite box size of our simulation, we limit our calculation $z<0.04$ so that we are not oversampling the box.  Unfortunately, there is a difference in box size between the MultiDark simulation and our simulations.  In order to fairly compare the bulk flow measurements, we randomly pick a box of the same size in the MultiDark simulation and then perform a similar calculation with 200 iterations.

In Figure 6, we compare two bulk flow measurements from the $11eV$ MOND simulations to observational \citet{kashlinsky} data, and the $10^{14}M_{\odot}$ mass cut bulk flows of the MultiDark simulation.  We find that these MOND data sets agree with the \citet{kashlinsky} data to within $1\sigma$; however, the trend for the MOND data points appears to be increasing slightly faster than the observational data.  The two bulk flow measurements from the  $11eV$ MOND simulations correspond to two different choices of the cluster cutoff radius ($1Mpc$ and $R_{200}$).  We see that these two data sets agree, which is direct evidence for our claim that the peculiar velocities are independent of our definition of the cluster despite the fact that we have computed the mass weighted bulk flow.  We find this agreement because the clusters that have a larger mass within $1Mpc$ will also likely have a greater mass within $R_{200}$.  As expected, the bulk flow measurements from the MultiDark simulation are far less than both the observational data and what was measured from the $11eV$ MOND simulations.

Because our simulations are run with periodic boundary conditions it is likely that we are overestimating the the real numerical bulk flows that are present in the simulation.  Even though we are choosing volumes of spheres that are much less than the size of the box, the Monte Carlo method applied will also choose centers which are close to the edge of the box.  Since we are sampling repeated patters, it is almost certain that the bulk velocities measured are over estimates of the actual bulk flows.  This effect is less likely to be prominent in our analysis of the Multidark simulation.  In order to compare to our MOND simulation, we have isolated a similar volume inside the Multidark simulation which is $\thicksim1/8$ smaller than the total volume of the Multidark simulation.  It is far less likely that we are sampling the periodicity at the edges of the box to the degree that we are in the $11eV$ MOND simulations, thus the values extracted are less of an over estimate than in the MOND simulation.  The degree to which we are over estimating the real values is unknown.

It should be noted that the actual magnitude of bulk flows remains a matter of active debate.  \cite{Keisler2009,Osborne2011,Mody2012,PlanckBulkFlow} find no significant bulk flows, consistent with \LCDM.  However \cite{atrio2010} argues that the \cite{Keisler2009} study suffered from flaws in the analysis and \cite{atrio2013} claims that \cite{Osborne2011,Mody2012} use a different filtering scheme from the \cite{kashlinsky} analysis which leads to contradictory results.  Furthermore \cite{atrio2013} questions the statistical significance of what was measured by \cite{PlanckBulkFlow}.  Regardless of the debate on the observational data, our $11eV$ MOND simulations predict a large magnitude bulk flow which differentiates it from $\Lambda$CDM simulations.

\section{Finding Bullet Clusters}
While many groups have attempted to extract Bullet Cluster type systems from large N-body simulations beginning with \citet{hayashi}, with more recent attempts including \citet{farrarandrosen} and \citet{leeandk}, there still remains a debate on the exact mass and velocity profile of the system.  The prominent bow shock that is quite evident in X-ray images of the system is estimated to correspond to a velocity of $\thicksim4700\;\kms$ \citep{markevitch2002, markevitch2006}.  However, using hydrodynamical simulations, \citet{springel2007} and \citet{milo2007} have estimated the velocity of the interacting subcluster to be $\thicksim2700\;\kms$ and $\thicksim4050\;\kms$ respectively, relative to the main cluster.  On the contrary, \citet{mastropietro2008} found that in order to best reproduce the structured X-ray profile of the cluster, the subcluster would need a velocity of $\thicksim3000\;\kms$.  All of these estimates depend on the initial velocities attributed to the subcluster.  Additionally, while the mass of the cluster is estimated at $2.8\times10^{15}M_{\odot}h^{-1}$, the mass ratio between the host and subcluster has yet to be agreed upon \citep{nusser2007}.  \citet{mastropietro2008} found that a mass ratio of $6:1$ for the host cluster to the subcluster best fits the X-ray data.  \citet{leeandk}, however, cite that these large scale mergers are extremely rare in simulations and choose to study clusters which have ratios of $\geq10:1$. 

Because the cluster masses in our $11eV$ MOND simulations tend to be on the order of $10^{15}M_{\odot}$ even at $z>0.3$ (the Bullet Cluster is located at $z=0.296$), we are able to limit the main cluster mass to only those greater than $10^{15}M_{\odot}$.  Because the box size of our $11eV$ MOND simulations is much smaller than both the MultiDark and \citet{leeandk} simulation, it remains difficult to determine an exact probability for finding a Bullet Cluster type object, however we are able to determine whether these objects are consistent with our simulations.  We identify possible Bullet Cluster type objects by measuring the probability of a head on collision.  We follow the methodology of \citet{leeandk} by selecting a pair of clusters if $|\vec{V_c}\cdot \vec{r}|/(|\vec{V_c}||\vec{r}|)\geq90\%$ where $V_c$ is the velocity of the subcluster.

Since the mass determinations for the clusters were measured based on the 1 Mpc core of the object, the mass ratios and distances between clusters are not exact measurements.  Therefore, when attempting to identify these clusters, we seek to understand whether the pairwise velocities needed to create the unique bow shock are present in the simulation.  Plotted in Figure 7 are the pairwise velocities against distances between the centers of the clusters with the mass ratio of the two clusters indicated by shape. 

The velocities necessary to obtain a bow shock are present in the simulation. Some pairwise velocities exceed what is necessary at that distance.  These high pairwise velocities suggest that a Bullet Cluster type object is consistent with our $11eV$ MOND simulations.

\begin{figure}[t]
\epsscale{1.2}
\plotone{\figname{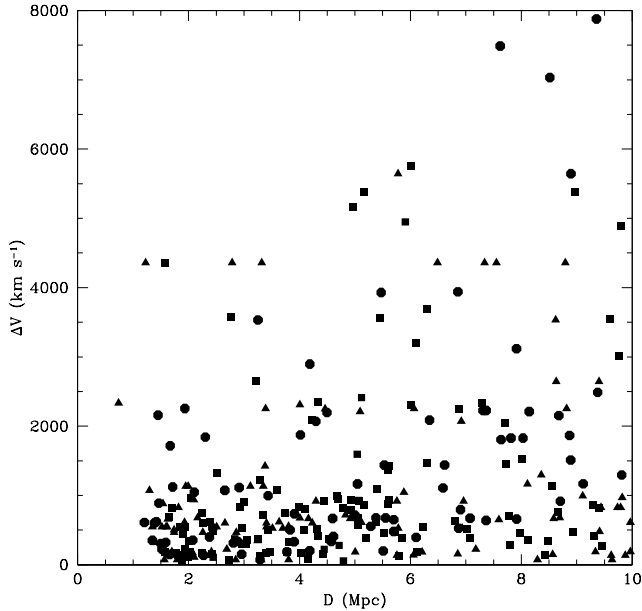}}
\caption{Pairwise velocity versus distance for Bullet type clusters identified in the eight $11eV$ MOND simulations.  Cluster pairs with small-to-large sub-cluster mass ratios $< 1:3$ are squares, circles are those pairs with $1:3 <$ mass ratio $< 2:3$ , and triangles have mass ratio $> 2:3$.  There is no apparent dependence of pairwise velocity on either separation or mass ratio.  High speed ($> 3000\kms$) collisions are not uncommon.}
\label{fig:}
\end{figure}

\begin{figure}[t]
\epsscale{1.2}
\plotone{\figname{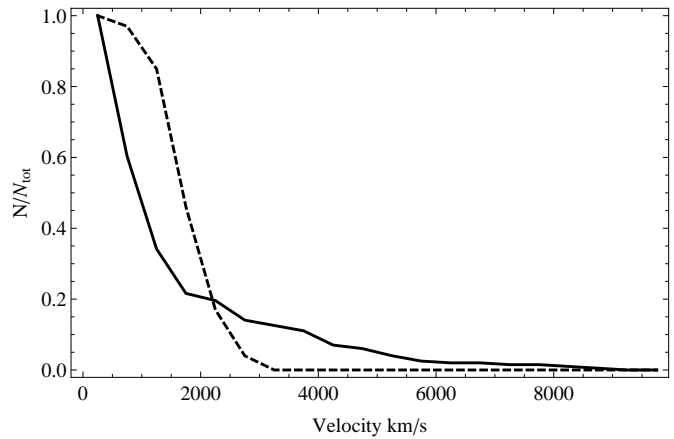}}
\caption{Cumulative distribution function of the clusters identified as Bullet Cluster candidates.  The candidates from the eight $11eV$ MOND simulations are shown as the solid black line and the candidates from the MultiDark simulation are shown as the dashed black line.  The CDF for eight $11eV$ MOND simulations extends well beyond that of the MultiDark simulation.}
\label{fig:}
\end{figure}

In order to recognize a potential Bullet Cluster type system in the MultiDark simulation, we employ similar criteria to \citet{leeandk}.  We search for satellite clusters within  $r<3M_{vir}$ of the main cluster and derive the probability of a direct head on collision with the main cluster by $|\vec{V_c}\cdot \vec{r}|/(|\vec{V_c}||\vec{r}|)$.  If this probability is greater than $90\%$, we identify these two clusters as a potential Bullet Cluster progenitor.

Similarly to \citet{leeandk}, we would like to understand the distributions of velocities of clusters that will undergo a direct head on collision.  Because the size of the box of the Multidark is only $1\ Gpc$ we do not expect there to be large numbers of $10^{15}M_{\odot}$ clusters as discussed in the previous sections.  Therefore it would be unfair to limit the probability calculation to only those clusters because they are extremely under represented.  For this reason, we focus on pairs of clusters which had a host cluster mass greater than $0.5\times10^{15}M_{\odot}$ and a mass ratio between $20:1$ and $3:1$.

It is important to note that we do not find any clusters over the $3000\;\kms$ threshold which has been argued to be the best velocity to reproduce the X-ray morphology of the Bullet Cluster.

The particle size in the MultiDark simulation is 27 times smaller than that in \citet{leeandk} simulation and \citet{leeandk} obtain their sample from a volume that is 9 times larger than the MultiDark simulation.  While they have more potential clusters, they do not see any pairwise velocities that are above $2000\;\kms$ while there are a few pairs of clusters in the MultiDark simulation that are above this velocity.  Additionally, we have to consider the difference in criteria we used to identify these clusters.  We have allowed for clusters whose mass ratios are greater than $10:1$ although we find that there is not a heavy dependence on where we cut off the mass ratio.  Because the probability distribution is weakly correlated with the mass ratio, we choose ours based on which has larger numbers of clusters in order to create a larger sample.  

\citet{leeandk} fit the distribution of pairwise velocities to a Gaussian function; however, the distributions of velocities may not necessarily be of this form.  Figure 8 shows a cumulative distribution function (CDF) of the pairwise velocities of the Bullet Cluster type objects identified in both the MultiDark simulation and our eight $11eV$ MOND simulations.  The CDF of the MultiDark simulation goes to zero around 3000km/s while the CDF for the $11eV$ MOND simulations remains above zero far beyond 3000 km/s.  Figure 8 clearly shows that the pairwise velocities in our $11eV$ MOND simulations are larger than what is found in the MultiDark simulation.

As we have stressed before, since the expansion history in MOND is unknown, we may have evolved the clusters too far.  This would naturally lead to higher pairwise velocities.  Our inability to reproduce the correct mass function will also play an important role in creating these higher velocities.   

\subsection{The El Gordo Cluster}
Recent observations of ACT-CL J0102-4915 (the El Gordo Cluster) \citep{elgordo} demonstrate two merging clusters with similar properties to the Bullet Cluster but resides at $z\thicksim0.870$.  The mass of the main cluster is estimated at $2\times10^{15}M_{\odot}$ and the pairwise velocity between the clusters in roughly 2300 \kms\  \citep{elgordo2}.  El Gordo is right at the edge of the maximum mass-redshift relation of \citet{mortonson}.  About half the observational uncertainty lies below their line and half above.  The odds of finding one such object in the entire observable universe are thus roughly 50:50.

In addition to its large mass at high redshift, it is also unlikely for El Gordo to have developed its high pairwise velocity \citep[cf.][]{angusmcgaugh,leeandk}.  Such a system is also rare in the $11eV$ MOND simulations.  While clusters of this mass exist at this redshift in our simulations, there were very few pairs of clusters which had a high enough velocity, main cluster mass, and collision probability to match the El Gordo cluster.  It did, however, occur in our relatively modest box size, so El Gordo would appear to be less unlikely in MOND than in \LCDM.

\section{Discussion \& Conclusion}
We have presented eight N-Body realizations of structure formation in the $11eV$ MOND cosmology hypothesized by \citet{angus2011}.  We find that the velocities in the MOND simulations as well as the masses of the largest clusters tend to be higher that those in a comparison $\Lambda$CDM simulation.  This potentially provides a natural explanation of the surprisingly large cluster bulk flow measured by \citet{kashlinsky, kashlinsky2010}.  We also find a much greater likelihood of producing the high collision speed observed in the Bullet Cluster in these MOND simulations than in \LCDM.  Our finding of high velocity clusters is consistent with the work of \cite{Llin2009} as we also find that the cumulative distribution function of $V/\sigma$ extends to $\sim3$.  Finally, clusters become more massive earlier in MOND than in \LCDM, a result that is potentially relevant to observations of massive clusters at high redshift (``pink elephants'') like El Gordo \citep{elgordo}.

While MOND appears promising for explaining these extremes, we have yet to recover the observed shape of the cluster mass function.  The limited resolution of available simulations precludes a detailed assessment of this important aspect of the observations.  However, progress is being made in this area and future MOND simulations that can reproduce the observed cluster mass function will surely differ from $\Lambda$CDM simulations in their predictions of bulk flows and Bullet Clusters.  

\section*{Acknowledgements}
We are grateful to Massimo Ricotti and Mia Bovill for discussions on the analysis of simulations.  This work has been supported in part by NSF grant AST 0908370.  Furthermore, we would like to thank the referee for his extraordinarily helpful suggestions and comments.

\bibliographystyle{./apj}
\bibliography{./HKatz2012revised}

\label{lastpage}
\end{document}